# Vanna-Volga Method for Normal Volatilities

Volodymyr Perederiy*

October 2018, rev. November 2020

## Abstract

Vanna-Volga is a popular method for the interpolation/extrapolation of volatility smiles. The technique is widely used in the FX markets context, due to its ability to consistently construct the entire Lognormal smile using only three Lognormal market quotes.

However, the derivation of the Vanna-Volga method itself is free of distributional assumptions. With this is mind, it is surprising there have been no attempts to apply the method to Normal volatilities (the current standard for interest rate markets).

We show how the method can be modified to build Normal volatility smiles. As it turns out, only minor modifications are required compared to the Lognormal case. Moreover, as the inversion of Normal volatilities from option prices is easier in the Normal case, the smile construction can occur at a machine-precision level using analytical formulae, making the approximations via Taylor-series unnecessary.

Apart from being based on practical and intuitive hedging arguments, the Vanna-Volga has further important advantages. In comparison to the Normal SABR model, the Vanna-Volga can easily fit both classical convex and atypical concave smiles (‘frowns’). Concave smile patterns are sometimes observed around ATM strikes in the interest rate markets, particularly in the situations of anticipated jumps (with an unclear outcome) in interest rates. Besides, concavity is often observed towards the lower/left end of the Normal volatility smiles of interest rates. At least in these situations, the Vanna-Volga can be expected to interpolate/extrapolate better than SABR.

* Perederiy Consulting (founder and consultant), PhD (Viadrina University, Germany)

research@perederiy-consulting.de, www.perederiy-consulting.de

# Introduction

Vanna-Volga (VV) is a popular and market-oriented technique for the modeling and interpolating/ extrapolating of volatility smiles. This technique has been mostly applied in the foreign exchange (FX) markets. This can be explained by the fact that the technique allows for a consistent deduction of the entire volatility smile given just three market quotes, and the FX markets happen to have three most liquid standard quotes (so-called ATM, risk reversal and butterfly).

Besides, the FX focus of the technique seems to be based on the fact that FX is one of the markets where Lognormal modeling is still deemed feasible. Although the final interpolation and approximation formulae in the VV methodology (see Castagna and Mercurio [2007]) were indeed derived for the Lognormal case, the methodology itself is free of distributional assumptions. It is surprising there has been little or no theoretical nor empirical work on applying the VV method to the Normal (Bachelier) volatilities. Recently, such models became particularly important for the modeling of interest rates, given the arrival of the negative interest rates.

In this paper, we reiterate the derivation of the VV method, and obtain its results for the Normal/Bachelier case using hedging arguments analogous to Castagna and Mercurio [2007]. In doing so, we retort to forward (rather than spot) hedging instruments, as these are more natural in the interest rate context. We investigate the volatility patterns generated by the technique and compare them to those resulting from the classical Normal SABR method.

# Bachelier vs. Black Model

## Pricing Formulae

In the Normal/Bachelier option pricing model, the forward price of the underlying is assumed to follow a driftless Brownian motion (in the T-forward measure):

$$dF_t = \sigma dW_t \qquad (1)$$

where $\sigma$ is the (constant) volatility. Then, the call option price with a strike $K$ and expiry $T$ can be shown to equal:

$$C = P(0,T)\left[(F-K)\Phi(d) + \sigma\sqrt{T}\phi(d)\right]$$
$$d = \frac{F-K}{\sigma\sqrt{T}} \qquad (2)$$

with $\phi$ signifying the density, and $\Phi$ the cumulative distribution function of the normal distribution, and $P(0,T)$ being the appropriate discount factor. We use the term 'moneyness' for the above $d$ term.

This Bachelier formula can be compared to the classical Black 76 formula for the Lognormal case, where the assumption of the underlying process is:

$$dF_t = \sigma F dW_t \qquad (3)$$

Here, the call option price is calculated as:

$$C = P(0,T)\left[F\Phi(d_1) - K\Phi(d_2)\right]$$
$$d_+ = \frac{\ln\frac{F}{K} + \frac{\sigma^2}{2}T}{\sigma\sqrt{T}} \qquad (4)$$
$$d_- = \frac{\ln\frac{F}{K} - \frac{\sigma^2}{2}T}{\sigma\sqrt{T}}$$

with the two moneyness terms $d_+$ and $d_-$.

## Option Greeks

We now derive the relevant Greeks (option sensitivities) for the Bachelier case, which will be required later[1]. The Delta and Vega can be shown (see e.g. Frankena [2016]) to equal:

**(Forward) Delta:**

$$\text{Delta}_{\text{f}} = \frac{\partial C}{\partial F} = P(0,T)\Phi(d) \tag{5}$$

**Vega:**

$$\Upsilon = \frac{\partial C}{\partial \sigma} = P(0,T)\sqrt{T}\phi(d) \tag{6}$$

From this, the relevant second-order derivatives can be derived as follows:

**Volga:**

First, we have:

$$\frac{\partial^2 C}{\partial \sigma^2} = \frac{\partial \Upsilon}{\partial \sigma} = P(0,T)\sqrt{T}\frac{\partial \phi(d)}{\partial \sigma}$$

and, making use of the derivative of normal density $d\phi(x)/dx = -x\phi(x)$ and applying the chain rule, we arrive at:

$$\frac{\partial^2 C}{\partial \sigma^2} = P(0,T)\sqrt{T}\left[\left(-d\phi(d)\right)\left(-\frac{F-K}{\sqrt{T}}\sigma^{-2}\right)\right] = \frac{d^2}{\sigma}P(0,T)\sqrt{T}\phi(d) = \frac{d^2}{\sigma}\Upsilon \tag{7}$$

**(Forward) Vanna:**

Analogously:

$$\frac{\partial^2 C}{\partial F d\sigma} = \frac{\partial \Upsilon}{\partial F} = P(0,T)\sqrt{T}\left[\left(-d\phi(d)\right)\left(\frac{1}{\sigma\sqrt{T}}\right)\right] = P(0,T)\frac{-d\phi(d)}{\sigma} = -\Upsilon\frac{d}{\sqrt{T}\sigma} \tag{8}$$

**(Forward) Gamma:**

$$\frac{\partial^2 C}{\partial F^2} = \frac{\partial \text{Delta}_{\text{f}}}{\partial F} = \frac{P(0,T)\phi(d)}{\sigma\sqrt{T}} = \frac{\Upsilon}{\sigma T} \tag{9}$$

We now compare the Normal Greeks derived above to the Lognormal Greeks from the classical Black 76 model[2] (second-order derivatives are expressed in terms of Vega $\Upsilon$):

[1] The Greeks derived here are for the forward (and not spot) price *F*. Corresponding Delta, Vanna and Gamma Greeks related to the spot (instead of forward) *S* can be easily obtained via multiplying the forward-related Greeks with the partial derivative (with repeated multiplication for Gamma) $\frac{\partial \text{F}}{\partial S} = \frac{\partial (Se^{cT})}{\partial S} = e^{cT}$ and substituting $F = Se^{cT}$, where $c$ stands for the cost of carry.

[2] See Xiong (2016) and Frankena (2016) for the derivation of the Lognormal (Black 76) Greeks.

| Greeks | Normal (Bachelier) | Lognormal (Black 76) |
|---|---|---|
| Moneyness | $d = \frac{F-K}{\sigma\sqrt{T}}$ | $d_+ = \frac{\ln\frac{F}{K} + \frac{\sigma^2}{2}T}{\sigma\sqrt{T}}$<br>$d_- = \frac{\ln\frac{F}{K} - \frac{\sigma^2}{2}T}{\sigma\sqrt{T}}$ |
| Delta, $\frac{\partial C}{\partial F}$ | $P(0,T)\Phi(d)$ | $P(0,T)\Phi(d_+)$ |
| Vega, $\frac{\partial C}{\partial \sigma}$ | $\Upsilon = P(0,T)\sqrt{T}\,\phi(d)$ | $\Upsilon = P(0,T)F\sqrt{T}\,\phi(d_+)$ |
| Gamma, $\frac{\partial^2 C}{\partial F^2}$ | $\frac{\Upsilon}{\sigma T}$ | $\frac{\Upsilon}{F^2\sigma T}$ |
| Volga, $\frac{\partial^2 C}{\partial \sigma^2}$ | $\Upsilon\frac{d^2}{\sigma}$ | $\Upsilon\frac{d_+ d_-}{\sigma}$ |
| Vanna, $\frac{\partial^2 C}{\partial F d\sigma}$ | $-\Upsilon\frac{d}{\sqrt{T}\sigma}$ | $-\Upsilon\frac{d_-}{F\sqrt{T}\sigma}$ |

There are obvious similarities between the Normal and Lognormal Greeks. In particular, when expressed in terms of the corresponding moneyness (as defined above), the expressions are actually identical apart from the inclusion of the forward price in the Lognormal case. Also, given a fixed Vega, the Vanna Greek is in both cases linearly proportional to the moneyness, and the Volga – to the squared moneyness.

# Vanna-Volga for Normal/Bachelier Volatilities

## Portfolio Construction

As in the case of the Lognormal VV method (see Castagna and Mercurio [2007]), we assume that the option price can be described with a flat (strike-independent) but stochastic implied volatility. Analogously, we construct a portfolio $\Pi$ consisting of:

- one long position in the call option under consideration with strike $K_0$ and value $C_0$ (according to the Bachelier model)
- a short position $\Delta$ in the delta-hedging asset $H$ and
- three short positions $\mathrm{w}_1, \mathrm{w}_2, \mathrm{w}_3$ in some pivot/benchmark call options with strikes $K_1, K_2, K_3$ and values $C_1, C_2, C_3$ (according to the Bachelier model).

Over a small time interval, by application of Ito's lemma to the function $C(F,\sigma,t)$, we can derive the following changes in the options' values (for i=0,1,2,3), based on a change in the forward price $dF$ and a change in the implied volatility $d\sigma$:

$$
\begin{aligned}
dC_i = {} & \frac{\partial C_i}{\partial F}dF + \frac{\partial C_i}{\partial t}dt + \frac{\partial C_i}{\partial \sigma}d\sigma \\
& + \frac{1}{2}\left(\frac{\partial^2 C_i}{\partial F^2}(dF)^2 + \frac{\partial^2 C_i}{\partial \sigma^2}(d\sigma)^2 + \frac{\partial^2 C_i}{\partial t^2}(dt)^2\right) \\
& + \left(\frac{\partial^2 C_i}{\partial F \partial t}(dFdt) + \frac{\partial^2 C_i}{\partial \sigma \partial t}(d\sigma dt) + \frac{\partial^2 C_i}{\partial F \partial \sigma}(dFd\sigma)\right)
\end{aligned}
\tag{10}
$$

Now, because $dF = \sigma dW$ in the Bachelier case, it holds:

$$
(dF)^2 = \sigma^2 (dW)^2
$$

So, making use of the following rules of stochastic calculus:

$d\sigma dt = 0$

$(dt)^2 = 0$

$(dW)^2 = dt$

$dWdt = 0$

$dFdt = \sigma dWdt = 0$

the expression (10) can be simplified to:

$$dC_i = \frac{\partial C_i}{\partial F} dF + \frac{\partial C_i}{\partial t} dt + \frac{\partial C_i}{\partial \sigma} d\sigma + \frac{1}{2}\left(\frac{\partial^2 C_i}{\partial F^2}\sigma^2 dt + \frac{\partial^2 C_i}{\partial \sigma^2}(d\sigma)^2\right) + \left(\frac{\partial^2 C_i}{\partial F \partial \sigma}(dFd\sigma)\right)$$

or:

$$dC_i = \frac{\partial C_i}{\partial F} dF + \left(\frac{\partial C_i}{\partial t} + \frac{1}{2}\frac{\partial^2 C_i}{\partial F^2}\sigma^2\right) dt + \frac{\partial C_i}{\partial \sigma} d\sigma + \frac{1}{2}\frac{\partial^2 C_i}{\partial \sigma^2}(d\sigma)^2 + \frac{\partial^2 C_i}{\partial F \partial \sigma}(dFd\sigma) \qquad (11)$$

As for the delta-hedging asset, analogously by Ito's lemma, but now assuming independence between the price of this asset and volatility, we have:

$$dH = \frac{\partial H}{\partial F} dF + \left(\frac{\partial H}{\partial t} + \frac{1}{2}\frac{\partial^2 H}{\partial F^2}\sigma^2\right) dt \qquad (12)$$

The change in the overall portfolio value over a small time interval is then:

$$d\Pi = dC_0 - \Delta\, \mathrm{dH} - \sum_{i=1}^{3} w_i\, dC_i \qquad (13)$$

where $dC_0$, $dC_1, dC_2, dC_3$ are defined according to (11) and $\mathrm{dH}$ is defined according to (12).

So far, the derivation has been effectively assumption-free, apart from using the Normal relationship $(dF)^2 = \sigma^2(dW)^2$, which only effects the second drift term in front of $dt$ in (11) and (12), resulting here in $\frac{1}{2}\frac{\partial^2 C_i}{\partial F^2}\sigma^2$ and $\frac{1}{2}\frac{\partial^2 H}{\partial F^2}\sigma^2$ correspondingly. In the Lognormal case, the relationship is $(dF)^2 = \sigma^2 F^2 (dW)^2$, leading therefore to the same results with the terms changed to $\frac{1}{2}\frac{\partial^2 C_i}{\partial F^2}\sigma^2 F^2$ and $\frac{1}{2}\frac{\partial^2 H}{\partial F^2}\sigma^2 F^2$.

## Risk Elimination

From (13), and using the Normal Greeks derived in the previous section, we can calculate the quantities $w_1$, $w_2$ and $w_3$ and $\Delta$ such that this makes $d\Pi$ insensitive to changes in the forward price $dF$ and to changes in volatility $d\sigma$.

The Delta risk of the portfolio (stemming from the terms in front of $dF$) can be eliminated, for arbitrary weights $w_i$, via setting the amount of the delta-hedging asset $\Delta$ such that

$$\frac{\partial C_0}{\partial F} - \Delta \frac{\partial H}{\partial F} - \sum_{i=1}^{3} w_i \frac{\partial C_i}{\partial F} = 0 \tag{14}$$

The Vega risk of the portfolio (stemming from terms in front of $d\sigma$) can be eliminated via enforcing the following restriction on the weights $w_i$:

$$\frac{\partial C_0}{\partial \sigma} - \sum_{i=1}^{3} w_i \frac{\partial C_i}{\partial \sigma} = \Upsilon_0 - \sum_{i=1}^{3} w_i \Upsilon_{\mathrm{i}} = 0$$

or:

$$\Upsilon_0 = \sum_{i=1}^{3} w_i \Upsilon_{\mathrm{i}} \tag{15}$$

where $\Upsilon_{\mathrm{i}}$ signifies the Vega Greek of the i-th option according to the Normal/Bachelier model.

The Vanna risk of the portfolio (stemming from the terms in front of $dFd\sigma$) can be eliminated via enforcing the following restriction on the weights $w_i$:

$$\frac{\partial^2 C_0}{\partial F \partial \sigma} - \sum_{i=1}^{3} w_i \frac{\partial^2 C_i}{\partial F \partial \sigma} = -\Upsilon_0 \frac{d_0}{\sqrt{T}\,\sigma} + \sum_{i=1}^{3} w_i \Upsilon_{\mathrm{i}} \frac{d_i}{\sqrt{T}\,\sigma} = 0$$

which is equivalent to

$$\Upsilon_0 d_0 = \sum_{i=1}^{3} w_i \Upsilon_{\mathrm{i}} d_i \tag{16}$$

where $d_i$ signifies the moneyness of the i-th option according to the Normal/Bachelier model.

The Volga risk of the portfolio (stemming from the terms in front of $(d\sigma)^2$) can be eliminated via enforcing the following restriction on the weights $w_i$:

$$\frac{\partial^2 C_0}{\partial \sigma^2} - \sum_{i=1}^{3} w_i \frac{\partial^2 C_i}{\partial \sigma^2} = \Upsilon_0 \frac{d_0^2}{\sigma} - \sum_{i=1}^{3} w_i \Upsilon_{\mathrm{i}} \frac{d_i^2}{\sigma} = 0$$

which is equivalent to

$$\Upsilon_0 d_0^2 = \sum_{i=1}^{3} w_i \Upsilon_{\mathrm{i}} d_i^2 \tag{17}$$

The drift term in front of $dt$ for the portfolio is:

$$-\Delta\left(\frac{\partial H}{\partial t} - \frac{1}{2}\sigma^2 \frac{\partial^2 H}{\partial F^2}\right) + \frac{\partial C_0}{\partial t} - \sum_{i=1}^{3} w_i \frac{\partial C_i}{\partial t} + \frac{1}{2}\frac{\partial^2 C_0}{\partial F^2}\sigma^2 - \sum_{i=1}^{3} w_i \frac{1}{2}\frac{\partial^2 C_0}{\partial F^2}\sigma^2$$

$$= -\Delta\left(\frac{\partial H}{\partial t} - \frac{1}{2}\sigma^2 \frac{\partial^2 H}{\partial F^2}\right) + \frac{\partial C_0}{\partial t} - \sum_{i=1}^{3} w_i \frac{\partial C_i}{\partial t} + \frac{1}{2}\frac{\Upsilon_0}{\sigma T}\sigma^2 - \sum_{i=1}^{3} w_i \frac{\Upsilon_{\mathrm{i}}}{\sigma T}\sigma^2$$

$$= -\Delta\left(\frac{\partial H}{\partial t} - \frac{1}{2}\sigma^2 \frac{\partial^2 H}{\partial F^2}\right) + \frac{\partial C_0}{\partial t} - \sum_{i=1}^{3} w_i \frac{\partial C_i}{\partial t} + \frac{1}{2}\frac{\sigma^2}{\sigma T}\left(\Upsilon_0 - \sum_{i=1}^{3} w_i \Upsilon_{\mathrm{i}}\right)$$

Because of the already implied restriction $\Upsilon_0 = \sum_{i=1}^{3} w_i \Upsilon_{\mathrm{i}}$, the term after the last $\frac{1}{2}$ also vanishes.

Thus, provided all constraints are imposed and fulfilled as described above, the change in portfolio value $d\Pi$ becomes (instantaneously) deterministic and free of the risks related to the underlying (forward) price and its volatility:

$$d\Pi = \left( -\Delta \left( \frac{\partial H}{\partial t} - \frac{1}{2}\sigma^2 \frac{\partial^2 H}{\partial F^2} \right) + \frac{\partial C_0}{\partial t} - \sum_{i=1}^{3} w_i \frac{\partial C_i}{\partial t} \right) dt \tag{18}$$

We summarize the weight constraints as a system of linear (in $w_i$) equations:

$$\begin{aligned} \Upsilon_1 w_1 + \Upsilon_2 w_2 + \Upsilon_3 w_3 &= \Upsilon_0 \\ \Upsilon_1 w_1 d_1^2 + \Upsilon_2 w_2 d_2^2 + \Upsilon_3 w_3 d_3^2 &= \Upsilon_0 d_0^2 \\ \Upsilon_1 w_1 d_1 \ + \Upsilon_2 w_2 d_2 \ + \Upsilon_3 w_3 d_3 \ &= \Upsilon_0 d_0 \end{aligned} \tag{19}$$

This system is, again, remarkably similar to that in the Lognormal case, with just both Lognormal moneyness terms $d_+$ and $d_-$ replaced by the Normal moneyness $d$.

The solution is analogously, by Cramer's rule:

$$\mathrm{w}_1 = \frac{\begin{vmatrix} \Upsilon_0 & \Upsilon_2 & \Upsilon_3 \\ \Upsilon_0 d_0 & \Upsilon_2 d_2 & \Upsilon_3 d_3 \\ \Upsilon_0 d_0^2 & \Upsilon_2 d_2^2 & \Upsilon_3 d_3^2 \end{vmatrix}}{\begin{vmatrix} \Upsilon_1 & \Upsilon_2 & \Upsilon_3 \\ \Upsilon_1 d_1 & \Upsilon_2 d_2 & \Upsilon_3 d_3 \\ \Upsilon_1 d_1^2 & \Upsilon_2 d_2^2 & \Upsilon_3 d_3^2 \end{vmatrix}}$$

$$\mathrm{w}_1 = \frac{\Upsilon_0 \Upsilon_2 \Upsilon_3 (d_2 - d_0)(d_3 - d_0)(d_3 - d_2)}{\Upsilon_1 \Upsilon_2 \Upsilon_3 (d_2 - d_1)(d_3 - d_1)(d_3 - d_2)}$$

which results in:

$$\mathrm{w}_1 = \frac{\Upsilon_0 (d_2 - d_0)(d_3 - d_0)}{\Upsilon_1 (d_2 - d_1)(d_3 - d_1)} \tag{20}$$

and, using the definition of the Normal moneyness:

$$w_1 = \frac{\Upsilon_0 (K_2 - K_0)(K_3 - K_0)}{\Upsilon_1 (K_2 - K_1)(K_3 - K_1)} \tag{21}$$

Again, this solution is very similar to the solution for the Lognormal case, where just the logarithms of the strikes appear instead of the strikes in (21) (see Castagna and Mercurio [2007]).

The solutions for $\mathrm{w}_2, \mathrm{w}_3$ are derived analogously, and are in the Bachelier case:

$$\begin{aligned} w_2 &= \frac{\Upsilon_0 (K_1 - K_0)(K_3 - K_0)}{\Upsilon_1 (K_1 - K_2)(K_3 - K_2)} \\ w_3 &= \frac{\Upsilon_0 (K_1 - K_0)(K_2 - K_0)}{\Upsilon_1 (K_1 - K_3)(K_2 - K_3)} \end{aligned} \tag{22}$$

## Choice of Delta-Hedging Instrument

Currently, the Normal/Bachelier model is mostly used for option pricing in the interest rate markets, with the underlying here being an interest rates index. As two examples: for caplets/floorlets the underlying is an IBOR interest rate, for swaptions - a swap rate. There is no spot market for such interest rates indices. The rates are not directly tradable assets and thus cannot serve as the delta-hedging instrument *H*.

However, there are tradable assets $H$ without vega risk ($\frac{\partial H}{\partial \sigma} = 0$) with their price linked to these rates and which can thus be used for the delta hedge. These are typically forward-like instruments whose current fair value is determined as:

$$H_t = P(t,T)\big(F_{t,T} - X\big)$$

where $F_{t,T}$ is the forward price/rate at time *t* for a delivery in *T* and $X$ is the strike (of the forward instrument). An example of such instrument is e.g. forward rate agreement (FRA), as the delta hedge for a caplet/floorlet option.

Generally, such forward-like instruments can be used as follows. By Ito's lemma, we have:

$$dH = \frac{\partial H}{\partial F} dF + \left(\frac{\partial H}{\partial t} + \frac{1}{2}\sigma^2 \frac{\partial^2 H}{\partial F^2}\right) dt$$

$$dH = P(t,T)\, dF + \left(\frac{\partial P(t,T)}{\partial t}\big(F_{t,T} - X\big) + 0\right) dt$$

Choosing a forward contract with zero fair value (i.e. with $X = F_{t,T}$) as the delta-hedging instrument, we obtain:

$$dH = P(t,T)\, dF$$

which can be substituted into (13) to obtain the quantity $\Delta$ of the delta-hedging instrument needed for the risk elimination.

## Smile Construction

So far, we have dealt with pricing in a world with a constant Bachelier/Normal volatility. However, the already mentioned value $\Pi$ of the resulting weighted portfolio in this world can be shown (see Shkolnikov [2009]) to be equal to the market value $\Pi_{\mathrm{mkt}}$ of the portfolio in the real world, with a stochastic (but strike-independent) implied volatility. As the value of the delta-hedging instrument *H* does not depend on volatility, we have:

$$C_{0,mkt} - \sum_{i-1}^{3} w_i\, C_{i,mkt} = C_0 - \sum_{i-1}^{3} w_i\, C_i \qquad (23)$$

The above relationship between the option market prices $C_{0,mkt}$ and $C_{i,mkt}$ can be transformed into a relationship between Normal volatilities $\sigma_0$, $\sigma_i$ implied by these prices. After rearranging

$$C_{0,mkt} - C_0 = \sum_{i-1}^{3} w_i\,\big(C_{i,mkt} - C_i\big) \qquad (24)$$

the differences in market vs. theoretical prices ($C_{0,mkt} - C_0$ and $C_{i,mkt} - C_i$) can be approximated based on the differences between the implied market volatilities and the reference flat volatility ($\sigma_0 - \sigma$ and $\sigma_i - \sigma$ ) using Taylor expansions. The first-order approximation (with derivation analogous to that in Castagna and Mercurio [2007] for the Lognormal case) results in the following expression for the implied market volatility of the option with a strike $K_0$:

$$\sigma_0(K_0\ ) \approx y_i\sigma_1 + y_2\sigma_2 + y_3\sigma_3 \qquad (25)$$

with (for Bachelier model):

$$y_1 = \frac{(K_2 - K_0)(K_3 - K_0)}{(K_2 - K_1)(K_3 - K_1)}$$
$$y_2 = \frac{(K_1 - K_0)(K_3 - K_0)}{(K_1 - K_2)(K_3 - K_2)}$$
$$y_3 = \frac{(K_1 - K_0)(K_2 - K_0)}{(K_1 - K_3)(K_2 - K_3)}$$
and $y_1 + y_2 + y_3 = 1$

Obviously, $\sigma_0(K_0)$ is a quadratic function of $K_0$ which intersects the points $(K_1, \sigma_1)$, $(K_2, \sigma_2)$ and $(K_3, \sigma_3)$.

The more precise second-order approximation results in the following quadratic equation:

$$\frac{{d_0}^2}{2\sigma}(\sigma_0 - \sigma)^2 + (\sigma_0 - \sigma) - (P + \frac{Q}{2\sigma}) \approx 0$$

with:

$$Q = \sum_{i-1}^{3} y_i {d_i}^2 (\sigma_i - \sigma)^2$$

$$P = -\sigma + \sum_{i-1}^{3} y_i \sigma_i$$

This quadratic equation can be easily solved for $\sigma_0 - \sigma$, which results in[3]:

$$\sigma_0(K_0) \approx \sigma + \frac{-\sigma + \sqrt{\sigma^2 + {d_0}^2(2\sigma P + Q)}}{{d_0}^2}$$

with:

$$d_0 = \frac{F - K_0}{\sigma \sqrt{T}}$$
$$Q = \sum_{i-1}^{3} y_i {d_i}^2 (\sigma_i - \sigma)^2 \qquad (26)$$
$$P = -\sigma + \sum_{i-1}^{3} y_i \sigma_i$$

Again, these solutions for Bachelier/Normal case are actually identical to the Lognormal case (as in Castagna and Mercurio [2007]), with strike prices in the former replacing the log-strikes in the latter, and moneyness terms redefined.

That said, the above Taylor-series approximations are actually not needed for the Bachelier/Normal model. Instead, the market price of the option with a strike price $K_0$ can be calculated directly as:

[3] For the particular case $F = K_0$ (or $d_0 = 0$), it immediately follows from the quadratic equation: $\sigma_0 = \sum_{i-1}^{3} y_i \sigma_i + \frac{\sum_{i-1}^{3} y_i {d_i}^2 (\sigma_i - \sigma)^2}{2\sigma}$

$$C_{0,mkt} = C_0 + \sum_{i-1}^{3} w_i \left(C_{i,mkt} - C_i\right) \tag{27}$$

and the corresponding Normal volatility $\sigma_0(K_0)$ can be simply backed out from this market price. This inversion is significantly easier for the Normal volatilities, with several analytical methods delivering machine precision results. E.g. one such method was proposed in Choi et al. [2007] and proceeds for a call price $C_{0,mkt}$ and the corresponding put price[4] $P_{0,mkt}$ (with the same expiry *T* and strike *K*) as follows[5]:

$$\sigma_0(K_0) = \sqrt{\frac{\pi}{2T}}\left(C_{0,mkt} + P_{0,mkt}\right)h(\eta) \tag{28}$$

$$\eta = \frac{(F - K_0)/(C_{0,mkt} + P_{0,mkt})}{\tanh^{-1}\left((F - K_0)/(C_{0,mkt} + P_{0,mkt})\right)}$$

$$h(\eta) = \sqrt{n}\frac{\sum_{k=0}^{7} a_k \eta^k}{\sum_{k=0}^{9} b_k \eta^k}$$

$a_0$=3.994961687345134 e-1
$a_1$=2.100960795068497 e+1
$a_2$=4.980340217855084 e+1
$a_3$=5.988761102690991e+2
$a_4$=1.848489695437094 e+3
$a_5$=6.106322407867059 e+3
$a_6$=2.493415285349361 e+4
$a_7$=1.266458051348246 e+4

$b_0$=1.000000000000000 e+0
$b_1$=4.990534153589422 e+1
$b_2$=3.093573936743112 e+1
$b_3$=1.495105008310999 e+3
$b_4$=1.323614537899738 e+3
$b_5$=1.598919697679745 e+4
$b_6$=2.392008891720782 e+4
$b_7$=3.608817108375034 e+3
$b_8$=-2.067719486400926 e+2
$b_9$=1.174240599306013 e+1

## Choice of the Reference Volatility

The application of the VV method requires (for the second-order approximation in (26) and the exact inversion as in (28)) the input of the reference (flat) volatility $\sigma$. Normally, either the ATM-implied volatility or the mid implied pivot volatility $\sigma_2$ (with $\sigma_1 < \sigma_2 < \sigma_3$) are used. There are, however, no solid theoretical underpinnings which would advocate such usage. This can be regarded as the main drawback of the Vanna-Volga method.

Instead, the reference volatility can be treated as yet another degree of freedom. In particular, for a reference volatility given, the standard VV methodology fits the smile to three arbitrary market pivotal quotes. Treating the reference volatility as yet another parameter would normally result in the VV methodology being able to fit a smile to an additional fourth market quote.

[4] The put price $P_{0,mkt}$ can be derived from the call price $C_{0,mkt}$ using the put-call parity $C_{0,mkt} - P_{0,mkt} = P(0,T)(F - K_0)$.

[5] In the particular case of ATM (for F=K), the implied Normal volatility can be inverted directly from (2) as $\sigma_{ATM,mkt} = \frac{2}{\sqrt{T}}\Phi^{-1}\left(\frac{1}{2}(C_{ATM}, mkt \,/\, F + 1)\right)$

# Comparison to Normal SABR

## Normal SABR Model

During the last decade, the SABR model has established itself as the market standard for the interpolation/extrapolation of interest rate volatility smiles. The technique is based on the forward rates $F_t$ modeled with stochastic instantaneous volatility $\alpha_t$ (see Hagan et al [2002] for details):

$$\begin{gathered} dF_t = \alpha_t {F_t}^{\beta} dW_t \\ d\alpha_t = \nu\, \alpha_t\, dZ_t \\ dW_t dZ_t = \rho dt \end{gathered} \tag{29}$$

with four parameters

$0 \leq \beta \leq 1,\ \ \nu \geq 0,\ \ -1 < \rho < 1$ and $\alpha_0 > 0.$

Setting $\beta = 0$ results in the so-called Normal SABR model, and allows modeling of negative forwards, similarly to the Normal/Bachelier framework. However, the SABR volatility dynamics is always assumed to be Lognormal. This differentiates this method from the VV technique, which does not make any assumptions as to the dynamics of the stochastic volatility.

The Normal implied volatility smile resulting from the Normal SABR model can be approximated as[6]:

$$\begin{aligned} \sigma_0(K_0) &= \alpha_0 \frac{\zeta}{x(\zeta)} \left( 1 + \frac{2 - 3\rho^2}{24} \nu^2 T \right) \\ \zeta &= \frac{\nu}{\alpha} (F - K_0) \\ x(\zeta) &= \log\left( \frac{\sqrt{1 - 2\rho\zeta + \zeta^2} - \rho + \zeta}{1 - \rho} \right) \end{aligned} \tag{30}$$

Generally, the $\alpha_0$ parameter controls the height of the smile, $\nu$ - its deepness, and the $\rho$ parameter - its skewness/asymmetry. Having three free parameters, the SABR smile in most situations can fit three market quotes exactly (similarly to the VV technique) and can then be used for inter- and extrapolation. However, due to the model design (in particular, the Lognormal volatility dynamics), the smiles resulting from SABR are always convex (or flat in the extreme case of $\nu = 0$). Thus, the method fails in the case of concave smiles (so-called ‘frowns’, where the maximum implied volatility is near ATM) which sometimes occur. As VV is free of assumptions as to the dynamics of the stochastic volatility, it can also be expected to deal well with such situations.

## Practical Examples

The examples below compare the smiles fitted via Normal VV vs. Normal SABR for some constellations of market data.

The first example refers to a classical situation with three pivots where ITM/OTM volatilities exceed the ATM volatility, with slight skewness. The VV smile is fitted using the exact inversion method (see (28)).

[6] See Frankena [2016]. The original derivation in Hagan et al [2002] contains a typo in the relevant section ‘A5.Special Cases’ for the particular case of the Normal implied volatility for Normal SABR.

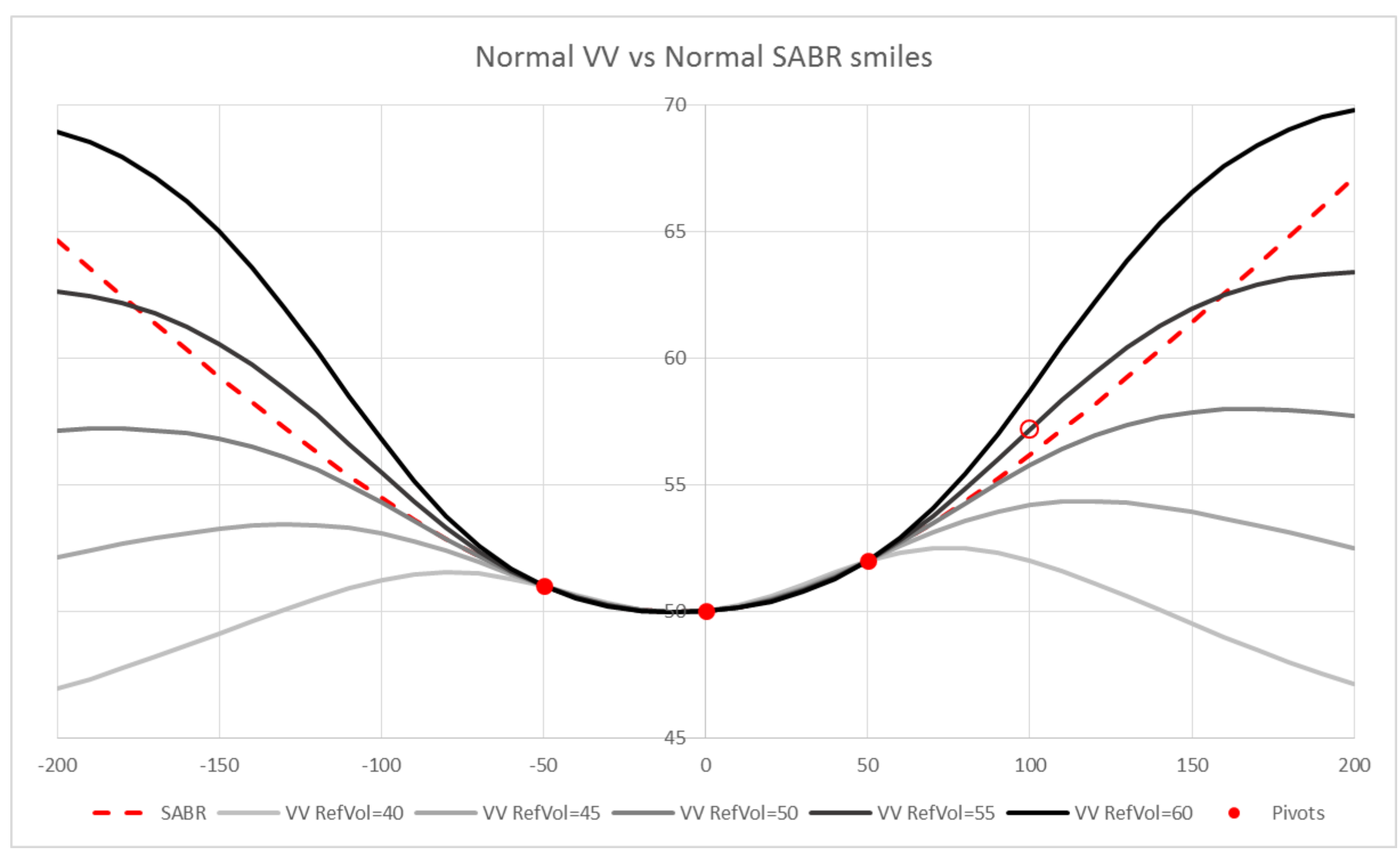

*Smile construction with Normal VV (with alternative reference volatilities of 40, 45, 50, 55, and 60) vs Normal SABR. For* $F = 0,\ T = 1,\ P(0,T) = 0,$ *and pivots* $\sigma_1(-50) = 51,\ \sigma_2(0) = 50,\ \sigma_3(50) = 52.$

As seen in this example, the patterns in the interpolation range are actually identical for VV vs. SABR, and also do not depend significantly on the VV reference volatility. The extrapolation patterns are quite different, however. In particular, the VV smiles show the "wings" on both sides, the smile becoming concave on the edges. The SABR smile remains convex, becoming almost linear on the edges. Furthermore, the VV smile extrapolation does depend significantly on the VV reference volatility: the higher it is, the higher the edges are. The closest match between the VV and SABR smile seems to occur with a VV reference volatility slightly exceeding the implied ATM volatility.

The strong dependence of the VV smile on the reference volatility advocates using a fourth pivotal quote, so that an exact fit is achieved for four pivots. In the graphical example above, this fourth quote is indicated by a circle at the strike 100, resulting in the selection of the optimal reference volatility equal to 55. Generally, inferring the optimal VV reference volatility via a fit to a fourth market quote can be achieved via simple numerical methods. Generally, such a four-pivot fit is impossible with SABR.

We now regard a less classical situation of an inverted smile ('frown'), where the ITM/OTM volatilities fall below the ATM volatility.

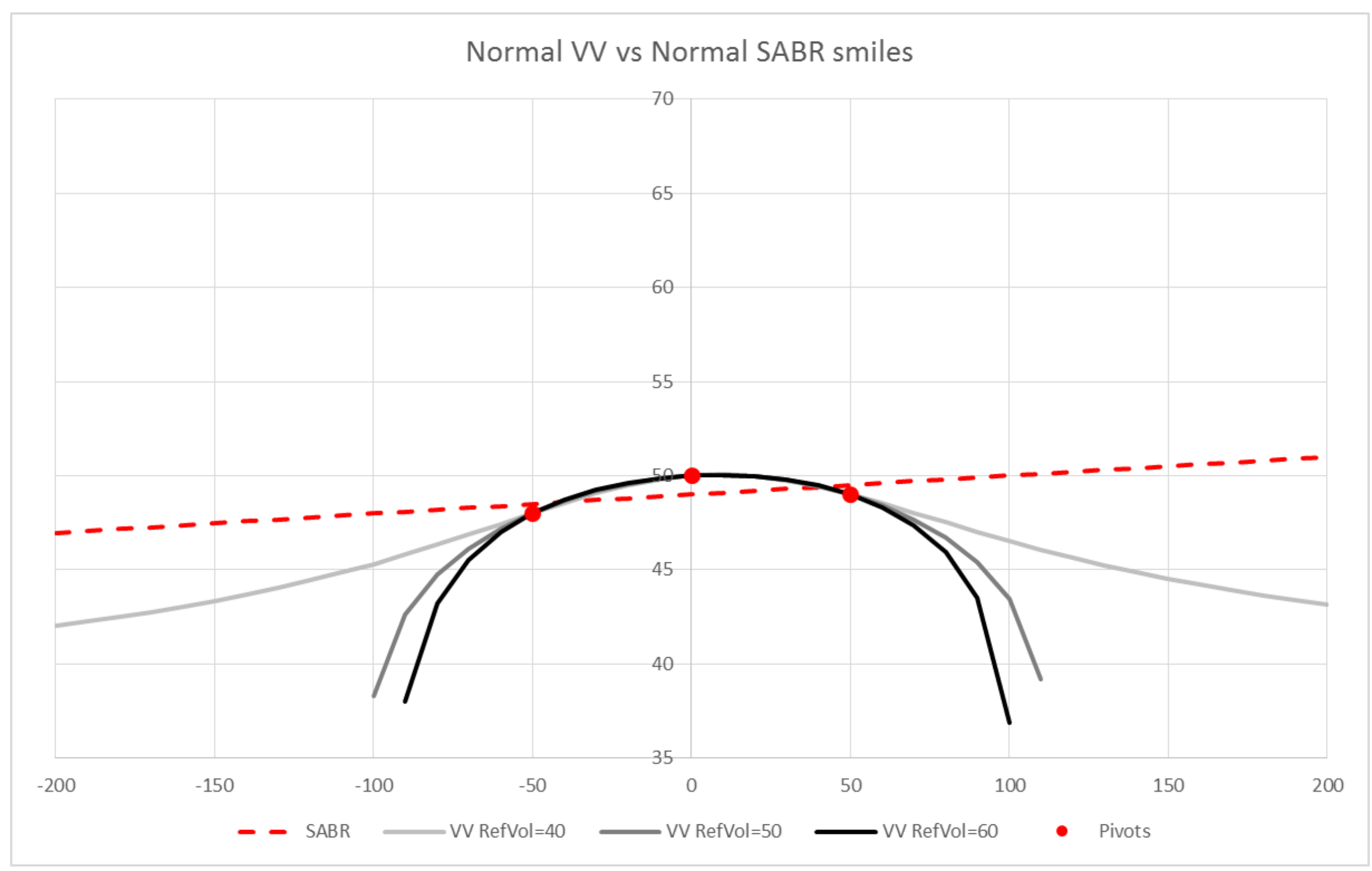

*Smile construction with Normal VV (with alternative reference volatilities of 40, 50, and 60) vs Normal SABR. For* $F = 0,\ T = 1,\ P(0,T) = 0,$ *and pivots* $\sigma_1(-50) = 48,\ \sigma_2(0) = 50,\ \sigma_3(50) = 49$

The best SABR fit is linear, as SABR is generally unable to fit concave smiles. The VV method interpolates the concave smile ('frown') well. However, the robustness of the VV extrapolation seems to depend on the reference volatility. For (relatively) high reference volatilities, the method fails on the edges. Technically, in these cases the calculated VV call prices (see (27)) fall below the intrinsic values of the options. For the reference volatilities well below the implied ATM volatility, the VV extrapolation seems to remain robust and consistent on the edges.

As a final check, we verify if the concave smiles fitted by VV produce positive risk-neutral probabilities. To this end, the density of the risk-neutral distribution of the price *x* of the underlying at expiry can be calculated from option prices using the second derivative of the option price with respect to the strike[7] (see Breeden and Litzenberger [1978]):

$$f(x) = P(0,T)\frac{\partial^2}{\partial K^2}\,Call(K) \tag{31}$$

We estimate this density via a discrete approximation using triplets of calls with strikes $x - \delta,\ x,\ x + \delta$ (for a small $\delta$) as:

$$f(x) = P(0,T)\frac{C_{mkt}(x+\delta) + C_{mkt}(x-\delta) - 2C_{mkt}(x)}{\delta^2} \tag{32}$$

where $C_{mkt}$ refers to the VV result (27).

The figures below show the densities for some selected concave volatility smiles (frowns).

[7] Alternatively, for a cumulative distribution function $F(x)$ and put prices: $F(x) = P(0,T)\frac{\partial^2}{\partial K^2}\,Put(K)$

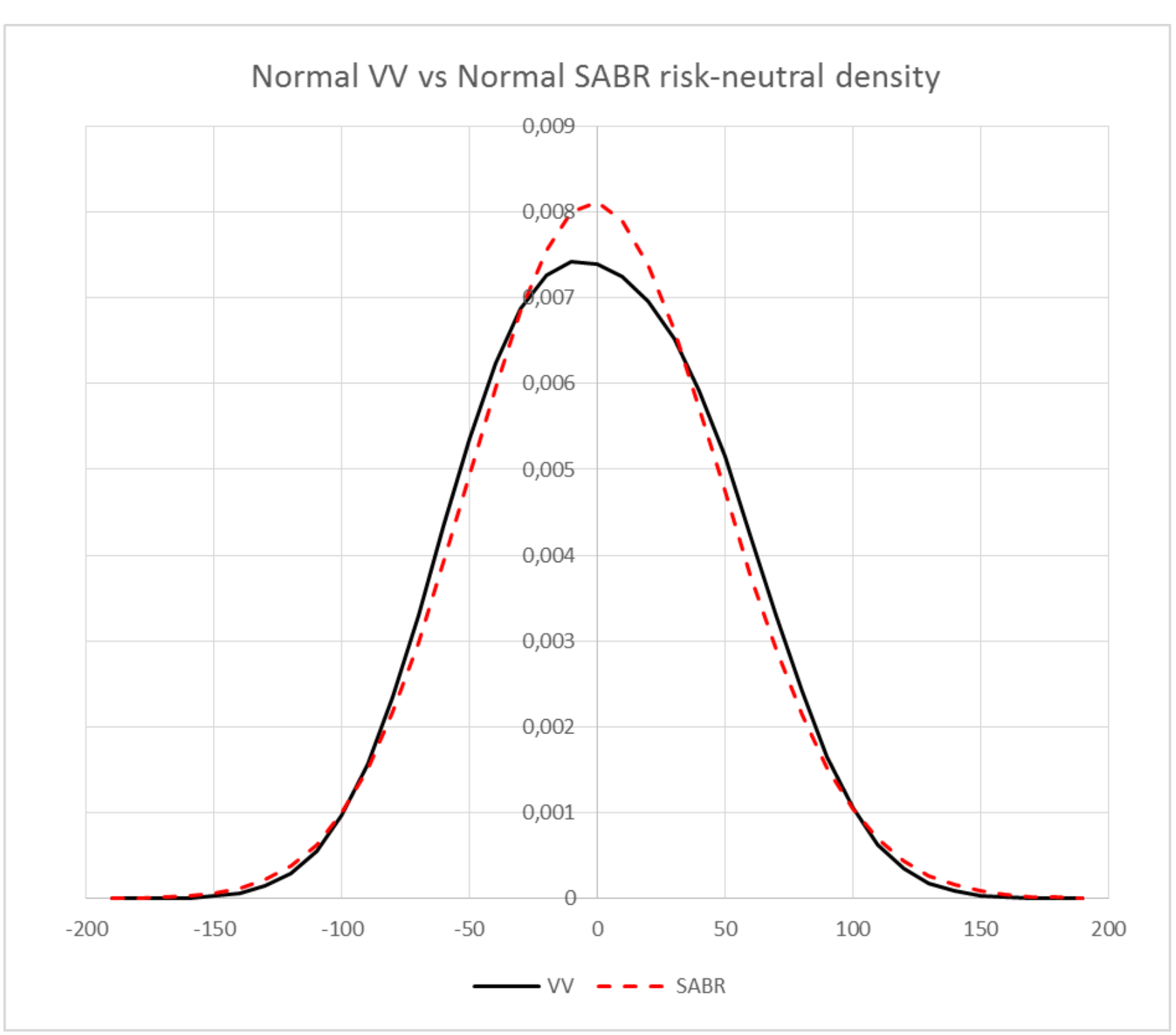

*Risk-neutral densities with Normal VV (for reference volatility=40) vs with Normal SABR. For* $F = 0,\ T = 1,\ P(0,T) = 0, and$ *pivots* $\sigma_1(-50) = 48,\ \sigma_2(0) = 50,$ $\sigma_3(50) = 49$

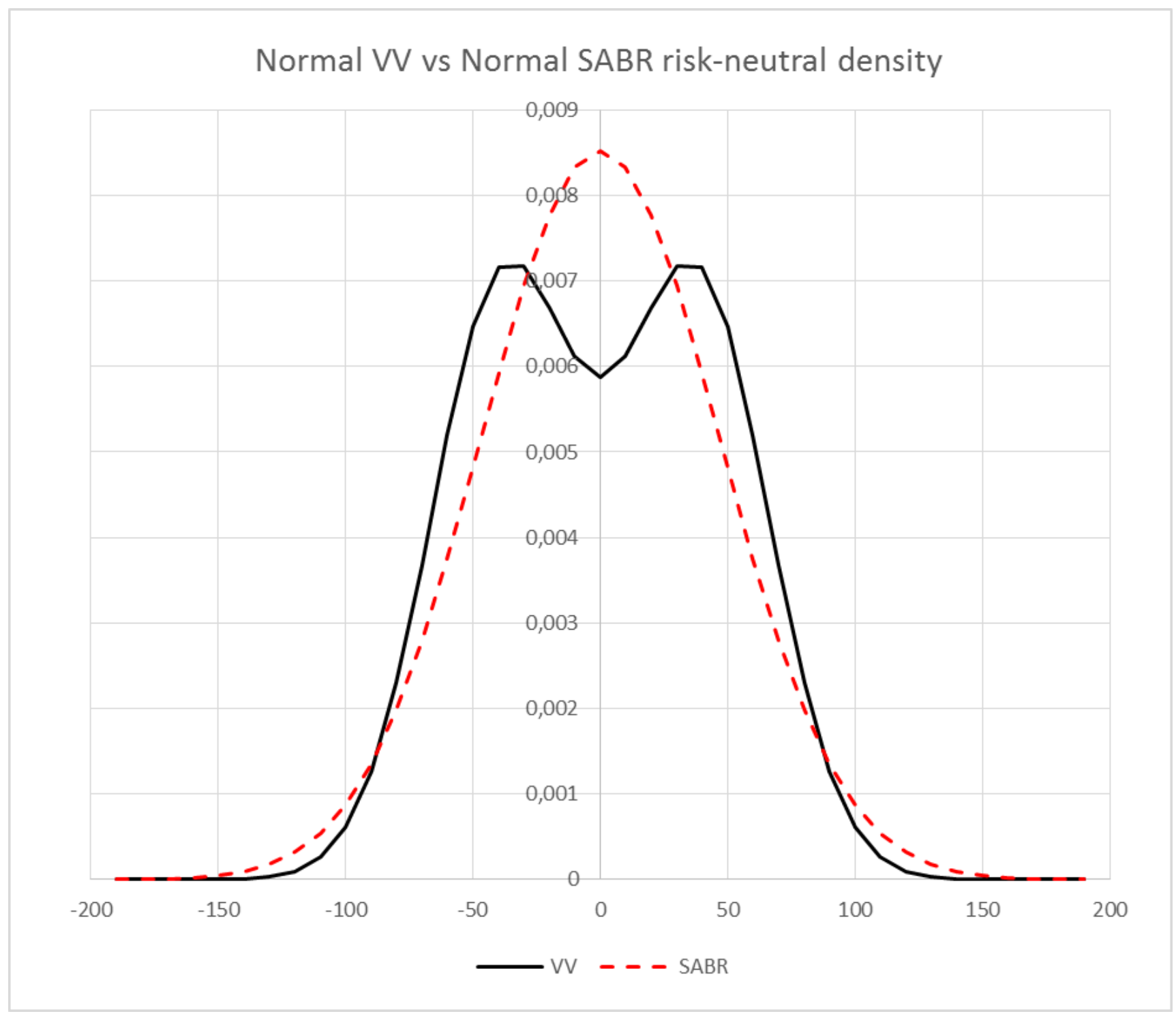

*Risk-neutral densities with Normal VV (for reference volatility=30) vs with Normal SABR. For* $F = 0,\ T = 1,\ P(0,T) = 0, and$ *pivots* $\sigma_1(-50) = 45,\ \sigma_2(0) = 50,$ $\sigma_3(50) = 45$

It becomes clear that even in the case of concave smiles (“frowns”) the VV technique can produce consistent non-negative risk-neutral densities. Moreover, with a frown deep enough, the density becomes bimodal, in line with expectations for situations with uncertain but significant jumps in the underlying price happening in the near future.

# Summary

We have shown that the Vanna-Volga technique is applicable for the Normal/Bachelier volatilities, with results quite similar to the classical Lognormal case, apart from different expressions for the option moneyness and for the Vega-related Greeks. The first-order Vanna-Volga approximation of the implied Normal volatility smile turns out to be just a quadratic function of the strike price. The more precise second-order approximation can also be easily derived for the Normal case and is similar to Lognormal. However, quasi-exact analytical solutions for the volatility smile can be easily derived in the Normal/Bachelier case due to the ease of the quasi-exact (machine-precision) analytical inference of implied volatilities from option prices, thus making the Taylor approximations unnecessary.

One of the drawbacks of the Vanna-Volga technique is its dependence on the reference volatility as one of the inputs. This volatility is generally unknown, but, as we show, can significantly influence the shape of the extrapolated smile. In practice, the ATM-implied volatility or a middle volatility is often used for this input, as a matter of intuition rather than logical reasoning. Instead, we propose treating the reference volatility as yet another degree of freedom. With the reference volatility becoming a model parameter, the Vanna-Volga methodology is generally able to fit a smile to four arbitrary market quotes.

The Vanna-Volga technique has some other important advantages. We compare the method to the Normal SABR model, which is the current interpolation/extrapolation standard in the interest rate context. The Vanna-Volga method seems to perform well in fitting the smile patterns typically observed in the interest rate markets. In contrast to SABR, the technique is able to fit the inverted (concave) smile patterns (or corresponding bimodal implied density patterns). These patterns are sometimes observed before important jump events (such as revisions of interest rates, with option maturities shortly thereafter). Besides, concavity is often observed towards the lower/left end of the Normal volatility smiles of interest rates, probably reflecting the strongly diminishing probability of interest rates becoming too negative. For these situations, the Normal Vanna-Volga technique can be expected to perform better than SABR.